\documentclass[12pt,preprint]{aastex}

\newcommand{\etal} {et~al.}
\def\spose#1{\hbox to 0pt{#1\hss}}
\newcommand\lsim{\mathrel{\spose{\lower 3pt\hbox{$\mathchar"218$}}
     \raise 2.0pt\hbox{$\mathchar"13C$}}}
\newcommand\gsim{\mathrel{\spose{\lower 3pt\hbox{$\mathchar"218$}}
     \raise 2.0pt\hbox{$\mathchar"13E$}}}
\newcommand{\agn}{{\small AGN}}
\newcommand{\blr}{{\small BLR}}
\newcommand{\nlr}{{\small NLR}}
\newcommand{\uta}{{\small UTA}}

\newcommand{\asca}{{\small \it ASCA}}
\newcommand{\ccd}{{\small CCD}}

\newcommand{\chandra}{{\it Chandra}}

\newcommand{\epic}{{\small EPIC}}
\newcommand{\om}{{\small OM}}

\newcommand{\fwhm}{{\small FWHM}}

\newcommand{\hetgs}{{\small HETGS}}
\newcommand{\hullac}{{\small HULLAC}}

\newcommand{\ngc}{{\small NGC}~3783}
\newcommand{\ngcx}{{\small NGC}}

\newcommand{\rgs}{{\small RGS}}
\newcommand{\rosat}{{\small \it ROSAT}}
\newcommand{\rrc}{{\small RRC}}

\newcommand{\usa}{{\small USA}}
\newcommand{\uv}{{\small UV}}
\newcommand{\xmm}{{\small \it XMM-Newton}}

\newcommand{\x}{X-ray}
\newcommand{\xs}{X-rays}
\newcommand{\cmcu}{cm$^{-3}$}
\newcommand{\cmsq}{cm$^{-2}$}
\newcommand{\kms}{km~s$^{-1}$}
\newcommand{\sm}{s$^{-1}$}

\begin{document}

\title{A Long Look at NGC 3783 with the \\
\emph{XMM-Newton} Reflection Grating Spectrometers }

\author{Ehud Behar\altaffilmark{1}, Andrew P. Rasmussen\altaffilmark{2},
  Alexander J. Blustin\altaffilmark{3},
%   Ari Laor\altaffilmark{1},
  Masao Sako\altaffilmark{4,5}, Steven M. Kahn\altaffilmark{2},
  Jelle S. Kaastra\altaffilmark{6},
  Graziella Branduardi-Raymont\altaffilmark{3},
  Katrien C. Steenbrugge\altaffilmark{6}}

\altaffiltext{1}{Physics Department,
  Technion, Haifa 32000,
  Israel; behar@physics.technion.ac.il}
  \altaffiltext{2}{Columbia Astrophysics Laboratory,
  Columbia University,
  550 West 120th Street,
  New York, NY 10027}
\altaffiltext{3} {MSSL, University College London, Holmbury St.
  Mary, Dorking, Surrey RH5 6NT, UK}
\altaffiltext{4} {Theoretical Astrophysics and Space Radiation
  Laboratory,  California Institute of Technology,  MC 130-33,
  Pasadena, CA 91125}
\altaffiltext{5}{\chandra\ fellow} \altaffiltext{6}{Space Research
  Organization of the Netherlands,
  Sorbonnelaan 2, 3548 CA, Utrecht, The Netherlands}

\begin{abstract}

A long 280~ks observation of the Seyfert 1 galaxy \ngc\ with \xmm\
is reported.  We focus on the oxygen line complex between 17 and
24 \AA\ as measured with the \rgs\ spectrometers. Accurate
absorption column densities and emission line fluxes are obtained.
We explore several options for the geometry and physical form of
the emitting and absorbing gas.  The lack of change in ionization
in the absorber despite an increase in continuum flux during the
observation restricts the high-ionization (O-K) and the
low-ionization (Fe-M) gas to distances of at least 0.5~pc and
2.8~pc, respectively, away from the central source. Given the
P-Cygni type profiles in the resonance spectral lines and the
similar velocity widths, column densities, and ionization
structure inferred separately from the emission and absorption
lines, it is tempting to relate the \x\ narrow-line emitting
plasma with the \x\ absorbing gas. Under this assumption, the
scenario of dense clumped clouds can be ruled out. Conversely,
extended ionization cones ($r \gtrsim$~10~pc) are consistent with
the observation independent of this assumption. These findings are
in stark contrast with the picture of numerous clumpy ($n_e
\gtrsim$ 10$^9$~\cmcu) clouds drawn recently from \uv\ spectra,
but it is consistent with the extended \x\ emission cones observed
directly in Seyfert~2 galaxies.

\end{abstract}


\section{Introduction}

As a bright, nearby Seyfert~1 galaxy [z = 0.00976, \citet {de91}],
\ngc\ is one of the best studied Active Galactic Nuclei (\agn). In
the \xs, observations with \rosat\ \citep {turner93} and \asca\
\citep {george98} revealed a highly-ionized absorber, but lacked
the spectral resolution required for kinematic measurements.
Lately, the rich absorption line spectrum of \ngc\ has been
readily measured with the grating spectrometers on board the \x\
observatories \chandra\ \citep {kaspi00, kaspi01, kaspi02} and
\xmm\ \citep {blustin02}, telling us that the highly-ionized gas
is outflowing at velocities of a few 100 \kms. These spectra of
\ngc\ feature numerous absorption lines by K-shell ions of C, N,
O, Ne, Mg, Si, and S, as well as by L-shell ions of Fe. Many
inner-shell absorption lines formed by lower ionization states are
also found in \ngc\ \citep {beharkabs02} indicating a broad range
of ionization.

\citet {kaspi02} used the high resolving power ($\lambda / \Delta
\lambda$ up to 1000) of the \chandra\ High Energy Transmission
Grating Spectrometer (\hetgs\, which covers the 1.6~- 23.4~\AA\
wavelength range) to determine the mean outflow velocity of $-$590
$\pm$ 150 \kms\ and a mean \fwhm\ of 820 $\pm$ 280 \kms\ for the
absorber in \ngc. The unprecedented small errors are due to an
integrated observing time of 900~ks. Both velocities are
consistent with a blend of known \uv\ absorption systems. \citet
{blustin02} employed a short 40~ks observation of \ngc\ with the
Reflection Grating Spectrometers (\rgs, $\lambda / \Delta \lambda$
up to 400) on board \xmm\ to construct a plasma model with two
ionization phases in the \x\ absorber that reproduced the \rgs\
spectrum between 6 and 38 \AA\ quite well. However, as was also
pointed out by \citet {kaspi01}, the wide range of observed charge
states is indicative of a broad distribution of ionization within
each phase.  Despite the extensive \x\ measurements, the location,
form, and energetic significance of the outflow in \ngc\ have
remained much of a mystery.

The connection of the \x\ absorber to \agn\ constituents observed
in other wavebands is a long standing puzzle.  The broad-line
region (\blr) in \ngc\ is inferred from the optical time lag
measurements of \citet {onken02} to cover a range of distances
from approximately 10$^{15}$ to 10$^{16}$~cm (not along the line
of sight). Combining \x\ and \uv\ observations, \citet {shields97}
favor the interpretation by which the \uv\ and \x\ absorbers
constitute a single medium. Their analysis implies an upper limit
of about 30~pc for the separation between the nucleus and the
absorbing plasma. Recently, four velocity components have been
resolved by \citet {gabel03} in the \uv\ absorption spectrum of
\ngc. Interpreting absorption of \ion {C}{3} at 1175~\AA\ as
arising from a metastable level, \citet {gabel03} estimate the
electron density in the high-velocity component to be $n_e \cong$
10$^9$ \cmcu\ and, hence, the absorber to be located at a distance
$r~\le$ 8~$\times$~10$^{17}$~cm away from the central source and
with a typical size of $\Delta r~\cong$ 10$^{10}$~cm, i.e.,
numerous highly-clumped cloudlets. Conversely, a narrow \ion
{O}{6} {\it emission} line component with a centroid velocity of
0~\kms\ was estimated by \citet {gabel03} to originate about 40~pc
away from the nucleus.

In December 2001, \xmm\ observed \ngc\ for 280~ks. The high
effective area of the \rgs\ makes this data set comparable in
statistical quality to that obtained in 900~ks with the \hetgs\ in
the respective wavebands. In this paper, we present the \rgs\
spectrum from this long observation. The main goal is to further
characterize the \x\ absorbing and emitting gas and particularly
to constrain its geometry and physical conditions.

\section{The Data}

\subsection{Data Reduction}

\ngc\ was observed by \xmm\ for a total of 280~ks from December 17
- 21 as part of the \rgs\ guaranteed time program.  The data
obtained with the \epic\ \x\ cameras and with the Optical Monitor
(\om) will be discussed elsewhere (Blustin et al., in
preparation). Here we focus on the \rgs\ spectrum. The \rgs\ data
were processed using custom software developed at Columbia
University, which operates on the observation data files (ODF).
This software is similar in origin and in function to the \rgs\
branch of the science analysis system (SAS), but is more flexible
and includes recent calibration of the instrumental O-edge region
based on accumulated data from long observations of Mrk~421 \citep
{andy03}.  It corrects for the effect of aspect drift on the
dispersion angles and includes background subtraction using
off-target \ccd\ regions.

\subsection{Short Time Variability}

The \rgs\ light curves are shown in Fig.~1. It is evident that the
soft \x\ flux varies fitfully and by as much as a factor of 2 on
time scales of hours. The light curves obtained with the EPIC
cameras, which are more sensitive to hard \xs, show very similar
behavior. The \uv\ light curve obtained with the Optical Monitor
(\om) resembles the curve in Fig.~1, only it is somewhat smoother
(Blustin \etal, in preparation). Observed variations in the
ionization state of the absorber due to changes in the continuum
flux could potentially shed light on the location of the absorber
through ionization and recombination times. With that goal in
mind, we divide the 280~ks observation into four temporal
segments, seeking variations in the absorption spectrum.

The four segments are labeled in Fig.~1 as the "low", "rise",
"high", and "peak" phases. The spectra obtained for each segment
separately are presented in Fig.~2. Each segment by itself suffers
from limited statistics. Consequently, disparities in narrow
features such as emission or absorption lines, which have limited
counts in a few bins, are very difficult to detect unambiguously.
On the other hand, it is clearly seen that the broad absorption
features in the spectrum, such as the Fe M-shell 2p~- 3d
unresolved transition array (\uta) between 16 and 17 \AA\ and the
\ion{O}{7} and \ion{O}{8} photoelectric bound-free K-edges (just
noticeable at 14.2~\& 16.8~\AA) do not vary significantly from one
time segment to the other. Indeed, calculated ratios of spectra
among the various time segments show mostly noise around the flat
ratio of the continua to within the data errors, i.e., $\sim$10\%.
For example, the ratio of the "high" and "low" segments is plotted
in Fig.~3. The plot is rather noisy, but no clear bound-free edge
features are observed. The only statistically significant
deviations from the mean in Fig.~3 are well localized around
strong emission/absorption lines (e.g., \ion{Ne}{9} at 13.48~\AA).
Since emission lines are not expected to disappear in the ratio
plot and since ratios at deep absorption troughs are unreliable,
we interpret the lack of bound-free features as evidence for no
detectable variation in absorption.

The position and shape of the Fe-M 2p-3d \uta\ are expected to be
particularly sensitive to the ionizing flux as several M-shell
charge states of the same element are observed at the same time.
During the present observation, the ionizing flux overall
increased gradually by almost a factor of 2 (Fig.~1). An average
increase of about 50\% occurs from the "low" state to the "high"
state. Such an increase in the ionization parameter $\xi$ [$\xi$~=
$L/(n_er^2)$, where $L$ is the ionizing luminosity] would have
shifted the \uta\ towards shorter wavelengths by about 0.4~\AA, if
the gas had reached a new equilibrium state \citep [c.f., Fig.~4
in ][] {beharuta01}.  Such a shift would have been surely
detectable with the \rgs.  Carefully examining Fig.~3, it is
possible to visualize a slight, statistically insignificant, shift
in the \uta\ around 16~\AA. However, this is exactly where the
\ion{O}{8} Ly$\beta$ line falls and in any event the effect is
smaller than would be expected from new equilibrium conditions.
Together with the lack of edge features in Fig.~3, this makes for
strong evidence that the ionization state in the absorber does not
have ample time to reach a new equilibrium state and, thus, it is
hardly affected by the rapidly varying continuum.

Since the ionization time depends directly on the absolute
ionizing flux seen by the absorber (i.e., its distance from the
source), we can use the lack of ionization despite the increase in
flux to place a lower bound to the separation between the source
and the absorber. For a detailed discussion of ionization in
winds, see \citet {krolik95}. We take the average power-law
ionizing flux density $F_E = A (E / E_0 )^{-\Gamma}$ with $E_0$ =
1~keV and $\Gamma \cong$ 1.7 \citep [determined for \ngc\ by] [and
here]{kaspi01, blustin02}. The normalization factor $A$ scales
with $r^{-2}$ and reaches its observed value of about 0.015
photons~\sm~\cmsq~keV$^{-1}$ at the telescope at a distance of
41.8~Mpc (assuming $H_0$~= 70 \kms~Mpc$^{-1}$). To first order and
as an upper limit, the equilibration time is the reciprocal of the
photoionization (PI) rate $\int{F_E\sigma ^{PI}(E)dE}$, where
$\sigma^{PI} (E)$ is the PI cross section. For \ion{O}{7}, we get
a time of 4.6~$r_{[pc]}^2$ days, where $r_{[pc]}$ is the distance
of the absorber from the source in parsecs. The unnoticeable
change in ionization observed over more than a day following the
rise of the continuum implies that the \ion{O}{7} absorber is at
least 0.46~pc away from the source. Extrapolating the power law
continuum down to the ionization energy of \ion{Fe}{9}
($\cong$~0.23~keV), which is dominant in the Fe-M UTA (at about
16.6~\AA), but also representative of the other M-shell ions, we
get an even shorter ionization time of about 0.13~$r_{[pc]}^2$
days. Consequently, the \ion{Fe}{9} absorber must be at least
2.8~pc away from the source to not be affected. Although only
crude lower limits, these estimates already clearly exclude the
\blr, which is approximately 0.003~pc away from the source \citep
{onken02}. For comparison with previous estimates, \citet
{appenzeller91} place emission by highly-ionized species (e.g.,
\ion{Fe}{7}) just outside the \blr, while \citet {shields97} give
an upper limit of 30~pc for the \x\ and \uv\ gas.

Note that despite the fluctuations over time scales of days, the
average \x\ flux in \ngc\ has remained steady over the past few
years \citep [e.g.,][]{markowitz03}. Under these conditions and at
a distance $\gtrsim$~0.5~pc, the ionization state of the absorber
does not react to the short time-scale fluctuations, but instead
it is governed entirely by the mean flux. Indeed, \citet {kaspi01}
point out that the shape of the \x\ absorption spectrum of \ngc\
is unchanged since 1996 when it was measured with \asca, although
the continuum level has clearly varied during that period. A study
of spectral variability on intermediate time scales of a few
months using fractions of the 900~ks \hetgs\ observation is bound
to shed new light on the behavior of \ngc\ in time (Netzer~\etal,
in preparation).

\subsection{Time Integrated Spectrum}

The fact that the absorption in the spectrum does not vary
significantly during the course of our observation justifies the
treatment of the 280~ks time integrated spectrum as a whole for
studying the absorption. Furthermore, when we compare the present
average spectrum with earlier observations \citep {kaspi01,
kaspi02, blustin02}, we still find no significant discrepancies.
Consequently, for the most part the best-fit model proposed by
\citet {blustin02} describes the current spectrum rather well.
This is demonstrated in Fig.~4 where the slab model from \citet
{blustin02} is compared with the current 280~ks spectrum. No
adjustment whatsoever has been applied to the model, including no
re-normalization.  Noticeable discrepancies between data and model
occur in the details around the O K$\alpha$ region between 14~and
24~\AA. These could not have been modeled properly with the first
40~ks exposure due to lack of counts in the narrow lines present
in this range.

In this work, we focus on the \ion{O}{1}~- \ion{O}{8} complex in
the wavelength range 17.3~- 24.7 \AA.  The goal is to measure
individual emission and absorption line fluxes. Due to
considerable blending, a simultaneous fit of many spectral
features is required.  We fit the spectrum locally with a
power-law plus an ensemble of individual-ion absorption models and
a few (unabsorbed) emission lines.  Subsequently, galactic
absorption ($N_H$~= 8.7$\times$10$^{20}$~\cmsq) is applied to the
entire spectrum.  The models are based on atomic data calculated
with \hullac\ \citep {bs01} and include all of the inner-shell
atomic transitions and their autoionization widths for all ions
and elements that appear in the spectrum. A similar method has
been successfully applied to several previous \agn\ spectra \citep
[e.g.,][]{sako01, kaastra02, blustin02}.

\subsubsection{Absorption}

Out of four velocity components resolved in the \uv\ (e.g., for
\ion{O}{6}), of which two are marginally resolved with \hetgs\ in
lines of \ion{O}{7} and  \ion{O}{8} \citep{kaspi02}, the \rgs\
reveals only one blended trough. Indeed, the blueshifts measured
here ($\cong~-$600 \kms) are consistent with the mean velocity of
the two dominant (\& slowest) components in the \uv\ and are in
agreement with the velocity measured with \hetgs. Thus, in the
model, we use only one velocity component for each ion.  We fit
for the turbulent velocity width and obtain $v_{turb}~= 170$~\kms\
(i.e., $\sigma_v$~= 120~\kms, \fwhm\ = 280~ \kms). This width is
very close to the value of 150~\kms\ used by \citet {kaspi00}.
Both here and in \citet {kaspi00}, this velocity width was used
because it enables the fitting of entire series of lines with a
single kinematic component. However, it should be clear that the
actual absorption troughs comprise more than one intrinsic
velocity component and are broader. In fact, \citet {kaspi01,
kaspi02} measured directly $v_{turb}$~= 500 $\pm$~170~\kms. In
fitting the spectrum, we assume a covering factor of unity, which
is an approximation for the value of 0.8 obtained by \citet
{gabel03}, who fit for the covering factor of the \x\ observed
components in the \uv\ spectra. The present velocity-blended
spectrum is not sensitive to this difference. Emission lines are
modeled by Gaussians. None of these approximations hamper the good
estimates we get for the total column density in each ion and for
the total flux in each emission line, which are the physical
quantities of interest here.

The best-fit spectral model for the O region is given in Figs.~5
and~6, highlighting respectively the K$\alpha$ and K$\beta$ (and
higher) transition regions. The good fit, including that for the
high-order lines of the He-like series below 19~\AA, is apparent
in the figures.  Absorption by all charge states from \ion{O}{8}
down to \ion{O}{4} is clearly detected.  It is obvious from the
figures that it is crucial to fit all charge states
simultaneously, including both absorption and emission, in order
to obtain the correct physical measurements. The various ions are
found outflowing at (indistinguishable) centroid velocities of
$-$470~- $-$800 \kms\ with respect to the systemic velocity of
\ngc. The fitted parameters for each of these ions are given in
Table~1.

Evidently, the broad range of charge states observed along the
line of sight is ubiquitous to Seyfert outflows \citep {sako01,
beharkabs02, blustin03, katrien03}. It is also observed in
Seyfert~2 galaxies \citep {ali02}. If all charge states are
present at similar distances from the central source, as observed
directly in Seyfert~2's, it implies a strong density gradient of
3-4 orders of magnitude (e.g., for \ion{O}{4} to \ion{O}{8}) over
relatively short distances. In turn, comparable column densities
together with a strong density gradient would imply that the
highly-ionized gas occupies a much larger volume than the
less-ionized gas. This picture is consistent with a wind that
emanates from a dense object (e.g., accretion disk or bloated
stars). It then accelerates, expands and becomes hotter and more
ionized. Pressure equilibrium also favors the hotter gas being
more tenuous and therefore more extended.

In addition to the high charge states, weak absorption by
\ion{O}{1} and \ion{O}{2} may also take place in (the rest frame
of) \ngc\ as indicated in Fig.~5. However, blending with other
ions hampers an unambiguous detection. The \ion{O}{1} K$\alpha$
line in \ngc\ overlaps with the strongest absorption lines of
\ion{Ar}{16} at 23.529~\AA\ \citep{jaan03}. The \ion{O}{2}
K$\alpha$ line in \ngc\ can not be resolved from the local
galactic K$\alpha$ line in our own galaxy. \ion{O}{3} is not
detected. Nevertheless, since we simultaneously fit many lines
from each ion, we are able to overcome the problem of blending to
some extent and to provide upper limits to the ionic column
densities. Of course, these constrains are not as tight as those
provided by \uv\ measurements for the same ions, when those are
available \citep {kraemer01}.

The errors on the outflow velocities are relatively large due to
the wavelength channel widths of the \rgs. Estimating the
uncertainty to be bound by 3 channels implies an error range of
$\pm$~16.5 m\AA\ corresponding at 22~\AA\ to $\pm$~230~\kms\ as
quoted in Table~1. The best-fit column densities are also given in
Table~1. The column densities of \ion {O}{7} and \ion {O}{8} are
constrained most tightly by their detected photoelectric edges. An
additional constraint is imposed by fitting entire series of lines
(see also Fig.~6). The values of (1.0 $\pm$~0.3) and (4.0
$\pm$~1.0)~$\times$~10$^{18}$~\cmsq\ obtained for the column
densities of \ion{O}{7} and \ion{O}{8}, respectively, are in
excellent agreement with the values of 1.1 and
4.3~$\times$~10$^{18}$~\cmsq\ obtained by \citet {kaspi02}.

The \ion{O}{6} ion is of particular interest for the possible
connection between the \uv\ and \x\ absorbers.  Inner-shell
\ion{O}{6} absorption can be observed in the present spectrum
although only marginally. We detect the K$\alpha$ line at a
significance of about 2 sigma over a total of 4 wavelength
channels, but the errors on the data points are still too large to
see the line profile clearly (compare data with continuum model in
Fig.~5). The feature at 19.33~\AA\ can be associated with the
\ion{O}{6} K$\beta$ line, although it is mostly due to \ion{O}{5}
K$\gamma$ (see Fig.~6). The derived column density of \ion{O}{5}
is much higher than that of \ion{O}{6}. This is due to the closed
shell structure of \ion{O}{5} and to the pervasively low
fractional abundance of the Li-like species \ion{O}{6}
\citep[$<$~0.25, e.g.][]{kallman01}. To that end, \ion{O}{6} and
Li-like ions in general are perhaps the least favorable vestige
for probing the ionized outflow in \agn. The precise restframe
wavelengths of the K$\alpha$ inner-shell transitions in O ions
have yet to be measured in the laboratory and are still somewhat
in debate \citep {beharlaw02}. Here, we use wavelength values for
\ion{O}{6} K$\alpha$ and K$\beta$ of 22.01 and 19.34 \AA,
respectively.  With a fitted turbulent velocity of 170~\kms, we
obtain only an upper limit to the \ion{O}{6} column density of 2.0
$\times$ 10$^{16}$~\cmsq. This is loosely consistent with \citet
{gabel03}, who find a lower limit of 0.63$\times$ 10$^{16}$~\cmsq\
for the total \ion{O}{6} column density in the two corresponding
\uv\ components. We note that there was a hint to the K$\alpha$
line of \ion{O}{6} in the 900~ks \hetgs\ spectrum \citep
{kaspi02}, but there, the lower statistical quality did not allow
for a definite identification.

The present high-quality spectrum reveals an absorption feature at
21.64~\AA\ (21.43~\AA\ in Fig.~5 at the rest frame of \ngc). We
identify it tentatively as absorption by \ion{O}{7} in the
intergalactic medium (IGM).  Given this interpretation, the gas
would be at a redshift of $z$~= 0.0018 $\pm$~0.0008. The
absorption parameters deduced for this line are given in Table~1.
Another possible identification for this line could be
\ion{Ca}{16} (21.49~\AA\ in the rest frame) in the outflow of
\ngc, but lines by neighboring charge states of Ca are not as
obvious in the spectrum, which casts doubt on the identification
of Ca. Since this feature may not be related to \ngc, we defer a
more careful analysis of it to a separate paper. There is a hint
of this line in the 900~ks \hetgs\ spectrum, but there it seems to
be at 21.60 \AA\ (as opposed to 21.43~\AA), where local \ion{O}{7}
would be expected. In any event, due to the lower statistical
quality of the \hetgs\ observation in this spectral band, \citet
{kaspi02} conclude that it is insignificant.

%\columnbreak
\subsubsection{Emission}

In addition to the numerous absorption lines, a few bright, narrow
emission lines are present in the spectrum.  In the O region of
the spectrum, the brightest emission lines are  \ion{O}{8}
Ly$\alpha$ at $\sim$~19~\AA\ and the He$\alpha$ triplet of
\ion{O}{7} at $\sim$~22~\AA\ (the resonance, intercombination, and
forbidden lines - $r, i, \&~f$). The radiative recombination
continuum (\rrc) of \ion{O}{7} is also detected, although its
location at 16.8 \AA\ in the midst of the Fe-M 2p~- 3d \uta\ makes
a precise determination of its flux rather difficult.  In fact, it
is much better resolved with \hetgs\ \citep [Fig.~6
in][]{kaspi02}. The weaker \ion{O}{7} and \ion{O}{8} K$\beta$ as
well as higher-order emission lines have been identified with
\hetgs\ \citep {kaspi02}, but here the lower spectral resolution
impedes the measurement of their flux. The measurable O emission
features are listed in Table~2. Fitting for a uniform velocity
width for all emission lines, we obtain a turbulent velocity of
740 $\pm$ 140~\kms\ (\fwhm\ 1220 $\pm$~230~\kms). \citet {kaspi02}
measured a \fwhm\ of 930 $\pm$ 350~\kms\ for the most isolated
\ion{O}{7} $f$-line. Previous optical measurements by \citet
{appenzeller88} of narrow, forbidden emission lines of
highly-ionized species in \ngc\ show a consistent linear increase
in velocity width with ionization energy. The presently measured
width of 1220~\kms~\fwhm\ for \ion{O}{7} fits that trend perfectly
(see their Fig.~13). For \ion{O}{8}, we are not able to provide an
independent measurement of the emission line width, because there
are no \ion{O}{8} non-absorbed (forbidden) lines in the spectrum.

As evident from Figs.~5 and~6, the emission/absorption model fits
the data (including the intricate line profiles) remarkably well.
This gives confidence in the individual (de-blended) line fluxes
obtained.  All of the measured emission lines are slightly
redshifted by about 300 $\pm$~230~\kms.  Owing to the large
errors, this is still consistent with the mean shift of 130
$\pm$~290~\kms\ measured for the emission lines with \hetgs\
\citep {kaspi02}.  These redshifts together with the complex
P-Cygni type profiles seen for the resonance lines are suggestive
of a bipolar outflow as discussed in detail below.  Also presented
in Table~2 are the line fluxes measured with \hetgs. The current
values are in good agreement with the \hetgs\ values. The largest
discrepancies are found for the resonance lines, which have
complicated P-Cygni type profiles [Ly$\alpha$ and
He$\alpha$($r$)]. This is due to the different methods used for
measuring the line fluxes. For example, \citet {kaspi02} did not
fit for overlapping absorption and did not require the widths of
the three \ion{O}{7} lines to be uniform. Nonetheless, the general
agreement between the two measurements implies no time variation
as expected for the narrow line region (\nlr).

\section{Discussion}

\subsection{Relation Between the X-Ray Emitter and Absorber}

To summarize the \x\ outflow kinematics in the rest frame of the
\agn\ as manifested by the absorption and emission lines: The
emission lines are redshifted by about 300~$\pm$~230~\kms\ and
have \fwhm\ of 1220~$\pm$~230~\kms\ (present work). The absorption
lines are blueshifted by about $-$600~$\pm$~150~\kms\ and have a
\fwhm\ of 820~$\pm$~280~\kms\ \citep{kaspi02}. Although it is
impossible to constrain the exact shape of the \x\ line profiles
and in particular their broad wings, due to the crowded nature of
this waveband and the overlapping emission and absorption, it is
obvious that high-velocity gas (up to $\sim$1000~\kms) is both
emitting and absorbing \xs. For the resonance lines, P-Cygni type
profiles are observed (Figs.~5 \& 6). These line profiles are
characteristic of a bipolar, non-collimated outflow in conjunction
with absorption by the outflow along the line of sight. It is
important to point out that we have only approximated these
profiles by emission and (slightly offset) absorption lines, but
the comparable velocity widths obtained for the absorption and
emission lines already supports the biconical interpretation.

We wish to further compare the \x\ line emission and absorption in
\ngc. In the context of Seyfert~2 galaxies, it has been shown that
intensity ratios among the He-like emission lines can be used to
decisively determine the column density through the emitting
plasma \citep {beharoutflows01, ali02} even when the absorption
spectrum is not available to us directly due to orientation
\&~obscuration effects. The present emission line fluxes give
ratios of $f / r$ = 1.5 $\pm$~0.3 and $i / r$ = 0.5 $\pm$~0.2.
With the velocity width of 740~\kms, this $f / r$ value is
obtained in our model at an \ion{O}{7} ionic column density of
(1.1 $\pm$~0.35) $\times$ 10$^{18}$~\cmsq. The $i / r$ ratio
provides just a lower limit of 1.3~$\times$~10$^{18}$~\cmsq. These
ranges are both in good agreement with the ionic column density of
(1.0 $\pm$~0.3) $\times$~10$^{18}$~\cmsq\ measured directly from
the absorption lines and edge of \ion{O}{7}.

Next, it would have been useful to compare the fractional ionic
abundances derived from the emission lines with those derived from
the absorption.  However, direct comparison is not available since
recombination-driven lines are due to the next ionized state
(e.g., \ion{O}{9} for Ly$\alpha$).  Hence, the \ion{O}{7} \&
\ion{O}{8} emission line intensities reflect the $n_{OVIII} /
n_{OIX}$ abundance ratio, while the absorption column densities
reflect $n_{OVII} / n_{OVIII}$. Nevertheless, the comparable
emission line fluxes and column densities of \ion{O}{7} and
\ion{O}{8} suggest that all highly-ionized states are fairly
abundant in the emitting plasma, just as they are in the absorbing
plasma. Moreover, it is possible to find an ionization parameter
$\xi$~=~30~erg~\sm~cm that reproduces just the right amount of
\ion{O}{7} and \ion{O}{8} emission and absorption, which we
observe. Clearly, we are seeing a very broad distribution in
$\xi$, but for simplicity, this single value can be used to
describe the \ion{O}{7} \& \ion{O}{8} spectral features. In
summary, the velocity widths, the column densities, and the
fractional abundances all make it tempting to interpret the \x\
emitter and absorber as arising from the same gas.

\subsection{Location and Geometry of the X-Ray Plasma}

The measured emission line flux $F_{ji}$ (e.g., of an O~ion) can
be related to the volume ($V$) emission measure $EM$ as follows:

\begin{equation}
EM \equiv \int_V{n_en_HdV}=\frac{4\pi d^2F_{ji}}{f_{q+1}A_OP_{ji}}
\end{equation}

\noindent The computed line power $P_{ji}$ is defined through the
line emissivity $n_e n_q P_{ji}$ (i.e., number of line photons
emitted in the plasma per unit time and per unit volume). Here,
$n_q$ is the number density of ions with charge $q$. The line
power takes into account all line-driving mechanisms and in
particular recombination, radiative cascades, and photoexcitation.
The fraction of ions with charge~$q$ ($n_q/\Sigma n_{q'}$) is
denoted by~$f_q$. For recombination-driven emission lines such as
the present ones, the most relevant fraction is $f_{q+1}$. In eq.
(1), $A_O$ represents the oxygen elemental abundance relative to
Hydrogen. The distance to the source $d$ is taken, for the case of
\ngc, to be 41.8~Mpc. In a fully ionized cosmic plasma, we can
also relate the electron density to the Hydrogen density by
$n_e$~= 1.2~$n_H$. On the other hand, the column density $N_H$ can
be written as follows:

\begin{equation}
N_H \equiv
\int_{r_{min}}^{r_{max}}{n_Hdr}=\frac{n_H}{n_e}\frac{L}{\xi}\left(
\frac{1}{r_{min}}- \frac{1}{r_{max}}\right)
\end{equation}

\noindent The measured ionic column densities $N_i$ are related to
$N_H$ by $N_i$=~$f_qA_ON_H$. The minimal and maximal distances for
absorbing gas in the outflow are denoted by $r_{min}$ and
$r_{max}$. To obtain the right hand side of eq.~(2), we used
$n_e$~= $L / (\xi r^2)$.

We focus here on the \ion{O}{7} and \ion{O}{8} spectral features
as the tracers of the very highly-ionized gas simply because they
exhibit the best measurable features in our spectrum. However,
since Seyfert~2 observations show that all charge states co-exist
and overlap in space along the outflow, we expect the present
analysis of O~ions to be applicable to the entire highly-ionized
plasma. We choose an ionization parameter of $\xi$~= 30~erg~\sm\
cm, where ionization balance calculations give $n_{OVIII} /
n_{OIX}\cong$~1.7 (in accord with the emission line ratio) and
$n_{OVIII} / n_{OVII}\cong$~4 (in accord with the absorption
column densities). This implies $f_q$ values of 0.14, 0.57, and
0.29, respectively for \ion{O}{7}, \ion{O}{8}, \& \ion{O}{9}.
Again, a single value of $\xi$ is assumed in this section just for
simplicity. The observed broad distribution in $\xi$ makes it
clear that each ion exists mostly around its typical range of
$\xi$.

Equations~(1) \&~(2) can not be solved for the most general case
and some approximations need to be made.  In the following we
explore the two extreme scenarios: $r_{max}~\cong r_{min}$ and
$r_{max}~\gg r_{min}$.

\subsubsection{Localized Scenario}

First, we explore the possibility of a well localized \x\ plasma,
i.e., one in which the typical size $\Delta r = r_{max} - r_{min}$
is much smaller than the average distance $r$ from the ionizing
source. In that case, the localization in $r$ immediately imposes
a narrow distribution in $n_e$ (through $\xi$).  Eqs. (1) and (2)
can then be rewritten simply as: $EM = n_Hn_e\Delta r^3$ and
$N_H=(n_H/n_e)(L/\xi)(\Delta r/r^2)$.  Even now, these equations
can not yet be solved separately. However, if one identifies the
emitter with the absorber, both equations can be solved
simultaneously. Indeed, by using eqs. (1), (2), and the definition
of $\xi$ to solve for $r$, $\Delta r$, and $n_e$ one would expect
to obtain a rough idea of these quantities. However, using the
present measurements for the \ion{O}{7} forbidden line and for the
\ion{O}{8} Ly$\alpha$ line, in both cases, we obtain values of $r
\cong \Delta r \cong$~10~pc. This suggests that the confined
geometry ($\Delta r~\ll ~r$) is inadequate for describing the
highly-ionized outflow of \ngc, at least as long as the line
emission and absorption are due to the same outflow constituent.
All of the measured and derived quantities are summarized in
Table~3.

\subsubsection{Ionization Cone Scenario}

Inspired by the \chandra\ \x\ images of Seyfert~2 galaxies
\citep{sako00, young01, brinkman02, ogle03}, we propose an
alternative geometry of two ionization cones with an opening solid
angle $\Omega$. In \ngcx ~1068, \agn\ driven \xs\ are detected as
far as 500~kpc away from the nucleus \citep {brinkman02}.
Furthermore, Young \etal\ (2001) \& \citet {ogle03} show that the
\x\ \nlr\ in \ngcx ~1068 overlaps on the plane of the sky with the
optical \nlr. Extended ionization cones in Seyfert~1's, only
oriented along the line of sight, would substantiate the unified
theory of \agn\ \citep{antonucci85} in the \x\ regime.
Unfortunately, it is impossible to distinguish between clumpy
clouds and ionization cones in \ngc\ by means of \x\ imaging, as
even the best angular resolution available with \chandra\
(0\arcsec .5) corresponds to 100~pc at the distance of \ngc.
Therefore, we have to make use of the spectrum.

In the ionization cones scenario, the plasma covers a large range
of distances (and densities) and it is necessary to revert to the
integral forms of $EM$ and $N_H$ in eqs. (1) and (2) with $dV =
2\Omega r^2dr$. Note that even within the framework of the
extended geometry and even when $r_{max} \gg\ r_{min}$, for a
given $\xi$ most of the absorption occurs close to $r_{min}$,
while the actual size of the cones (i.e., $r_{max}$) can not be
well constrained by the observation. Indeed, eqs. (1) and (2) can
now be rewritten as:

\begin{equation}
r_{min} \cong (r_{min}^{-1} - r_{max}^{-1})^{-1} = \frac{n_H}{
n_e} \frac{L}{\xi N_H}
\end{equation}

and

\begin{equation}
\Omega = \frac{1}{2} \frac{EM}{N_H} \frac{\xi}{L}
\end{equation}

In this scenario, since eq. (3) is attained from the column
density alone without the emission measure, the result for
$r_{min}$ is independent of the emission. Thus, it is independent
of the assumption that the \x\ emitter and absorber represent the
same plasma. Eqs.~(3) \&~(4) demonstrate the dependence of the
derived values on the assumption for $\xi$. Thus, if \ion{O}{7}
actually forms at $\xi$-values somewhat lower than we assume
(30~erg~\sm~cm), we would be slightly underestimating $r_{min}$
and overestimating $\Omega$, and vice versa. We stress that the
focus on a certain ion or on a specific value of $\xi$ constitutes
a complementary and totally consistent approach with the
observation of a broad range of charge states. Again, analogous to
the Seyfert~2 case, each charge state could be present throughout
the cone, only at different densities, while at the same time
different charge states are co-located thanks to the steep density
gradients at each point in the cone. The \x\ observation simply
traces \ion{O}{7} and \ion{O}{8} wherever they are along the cone.

We now substitute the measured values into eqs. (3) \& (4) using
the data for the \ion{O}{7} forbidden line and for the \ion{O}{8}
Ly$\alpha$ line. These values as well as the results for $r_{min}$
and for $\Omega$ are summarized in Table~3. The electron density
is again $n_e = L / (\xi r^2)$, but here it decreases rapidly
($\propto~r^{-2}$) along the cone. In Table~3 we give the maximal
$n_e$ value attained at $r_{min}$. It can be seen that the results
for both \ion{O}{7} and for \ion{O}{8} suggest consistently that
the \x\ emitter/absorber lies at $r_{min} \cong$~10~pc. The
maximal densities are of the order of 600 \cmcu. These results are
not very different from those obtained by \citet {shields97}, who
do not assume a cone geometry, but give values of $n_e
\gtrsim$~50~\cmcu\ and $r_{min} \lesssim$~30~pc.

The typical opening angle $\Omega$/2$\pi$ we get for each of the
two cones is about 15~- 25~\%, fully consistent with the fraction
of Seyfert galaxies that show a highly-ionized absorbing outflow
($\sim$50\% of the Seyfert~1's). The presently proposed geometry
with the high $r_{min}$ and $\Omega$ values makes for a large
scale outflow that from our perspective would entirely blanket the
central \agn\ region including the \blr, provided that the
ionization cone is uniformly filled. Since \citet {gabel03} find
that at some radial velocities the \blr\ is only partially covered
by the \uv\ outflow, identifying the \uv\ absorber with the
extended \x\ ionization cone would require that some of the \blr\
be observed directly, i.e., through (velocity-specific) holes in
the cone.  We note that \citet {katrien03} estimate for \ngcx
~5548 much smaller opening angles, $\Omega$/2$\pi~<$~0.5\%.
However, that result was obtained by assuming a mass loss rate
that is equal to the mass accretion rate (with a luminosity
efficiency of 5\%). If this assumption is relaxed, extended
ionization cones may be applicable for \ngcx ~5548 as well.
Indeed, the high values for $\Omega$ in \ngc\ imply that the
outflow mass loss rate exceed the accretion rate as discussed in
the following section.

\subsubsection{Mass Estimates}

By integrating the mass density $\mu n_Hm_H$ (where $\mu$~=~1.3 is
the average atomic weight for a solar-abundance plasma) over the
ionization cone (or clouds), it is possible to estimate the total
mass $M$ and similarly the mass loss rate $\dot{M}$ in the
outflow. These quantities are listed in Table~3 for both the
localized and ionization cone models.  In the ionization cone
model, the total mass depends on $r_{max}-r_{min}$, which is not
well constrained by the observation.  Therefore, the mass in
Table~3 is normalized to $r_{max} \cong r_{max}-r_{min}$. Taking
an outflow velocity of $-$600~\kms, we obtain a total mass loss
rate of up to a few solar masses per year. We compare this with an
estimate of the mass accretion rate. With a bolometric luminosity
of 4.5$\times$10$^{44}$~erg \sm\ \citep {markowitz03} and a
gravitation-to-light conversion efficiency of $L_{bol} /
(\dot{M}_{acc}c^2$)~=~5\%, \ngc\ would be accreting about 0.2
solar masses per year. If all these numbers are correct, then
\ngc\ is accreting only about 10\% of the mass it is ejecting.
Although this is a surprisingly low accretion efficiency, we are
not aware of any hard evidence against it. Furthermore, ultra-high
velocity outflows ($\sim$0.1c) have been discovered recently in
several \agn. It has been postulated that the high mass loss rate
observed in these sources is due to their high accretion rate
$L_{bol}/L_{Eddington}$ \citep {pounds03}. Narrow line Seyfert 1's
such as \ngc\ could be the low-mass facet of this phenomenon.

It is worth stressing that the errors given in Table~3 reflect
only those due to the presently measured $N_i$ \& $F_{ji}$ values.
Indeed, the present mass estimates should be treated with more
caution as the \x\ outflow of \ngc\ is undoubtedly more complex
than a simple, continuous cone, which was assumed here.
Nonetheless, these estimates are good enough to rule out radically
different scenarios, such as the highly-localized one. Most
likely, the true picture for \ngc\ is somewhat less extreme than
$r_{max}~\gg r_{min}$ and therefore $r_{min}$ could be somewhat
less than 10~pc, but certainly not much less, given the timing
constrains obtained in Sec.~2.2 of $r_{min}~\ge$ 0.5~pc for O-K
and $r_{min}~\ge$ 2.8~pc for Fe-M.

\subsubsection{Comparison With Other Observations}

The extended ionization cones stand in contrast with
interpretations of two other Seyfert~1 galaxies.  \citet
{netzer02}, based on observed recombination times, place the
ionized absorber in \ngcx ~3516 about 0.2~pc away from the
nucleus. \citet {kriss03} and \citet {blustin03} even suggest that
part of the ionized absorber in \ngcx ~7469 is closer to the
nucleus than the \blr.  In both of these cases, the observers find
that the \uv\ and \x\ absorbers are connected. More importantly,
the present results are in stark contrast with the parameters
derived by \citet {gabel03} for \ngc\ based on \uv\ absorption,
which according to those authors occurs at sub-parsec distances in
tiny clouds the size of $\Delta r \cong$~10$^{10}$~cm.  At the
heart of the argument of \citet {gabel03} lies the absorption by
the \ion{C}{3} multiplet at 1175 \AA\ detected in the
high-velocity \uv\ component (-1300~\kms). Following \citet
{bromage85}, \citet {gabel03} take this feature as evidence for
high-density gas ($\gtrsim$~10$^9$~\cmcu) in this component, which
inevitably restricts it to clumps very close to the source. We
wish to point out that the 2s2p triplet complex from which those
\ion{C}{3} lines are absorbed, can in fact be populated at lower
densities. Only one fine-structure level in the triplet requires
high densities \citep [see, e.g., Fig. 1 in][]{bhatia93}, but
absorption from the other two levels of the triplet may be
possible at low densities and might be producing the \ion{C}{3}
lines at 1175~\AA\ observed by \citet {gabel03}.

On the other hand, it may also be possible that dense clouds
reside at the base of the ionization cone where the outflow is
clumpy, less ionized, and thus absorbs \uv\ light.  Further out in
the outflow, perhaps, the ejecta evaporate, become much more
tenuous and consequently more highly-ionized.  These more extended
parts of the outflow might then constitute the \x\
absorbing/emitting gas. However, following \citet {kaspi02} and
\citet {gabel03}, we note that the very similar velocity structure
in the \x\ and \uv\ absorption systems strongly suggests that the
two coincide and most likely are two different manifestations of
the same outflow.  Finally, based on \uv\ variability and
recombination times, \citet {shields97} also arrive at the
conclusion that the \x\ and \uv\ absorbers are more than 10~pc
away from the ionizing source.

A paramount difference between the \uv -derived and \x -derived
parameters of the outflow lies in the corresponding $M$ and
$\dot{M}$ values.  In the \citet {gabel03} picture, the calculated
$M$ and $\dot{M}$ values are more than 10 orders of magnitude
smaller than the values derived here from the \x\ measurement (see
Table~3). According to the \uv\ interpretation, the outflow
carries insignificant mass and energy out of the \agn\ system,
while according to the present \x\ analysis, the outflow must play
a major role in the mass and energy budgets of the \agn.

\section{Conclusions}

The present 280~ks observation of \ngc\ leads to a few interesting
conclusions, which can be summarized as follows:

\begin{enumerate}

\item  Despite the continuum increasing by up to a factor of 2
during the 280~ks observation, no significant effect on the
absorption structure has been detected.  The fact that the
absorbing plasma did not respond to the increase in flux within,
say, 1~day places it at least 0.5~pc (O-K) to 2.8~pc (Fe-M) away
from the continuum source, regardless of its geometry or physical
situation (e.g., density).

\item The interpretation of the \x\ line absorbing and emitting
plasma being closely connected, or actually being manifestations
of the same outflow, is very tempting given the P-Cygni type
profiles as well as the similar velocity widths, inferred column
densities, and ionization distributions.

\item  Assuming that the two components are connected rules out
the localized geometry ($\Delta r~\ll~r$ and $n_e \cong
10^9$~\cmcu) for the \x\ plasma. This geometry was suggested by
\citet {gabel03} for the \uv\ absorbing outflow.

\item  Instead, we propose diffuse gas in the form of ionization cones.
These cones may not be closer than about 10~pc from the ionizing
source and must therefore have much lower densities (a few 100
\cmcu\ and decreasing with distance). In this case, we find that
the mass loss rate in the outflow is significant, of the order of
a solar mass per year.

\item  The conflict with the clumpy cloudlet description of \citet {gabel03}
could possibly be due to erroneous use of the \ion{C}{3} multiplet
at 1175~\AA\ as a density diagnostic, but it could also be
resolved if the \uv\ absorber lies at the base of the ionization
cone in the form of clouds. This constituent then represents a
transient phase of the gas as it expands, ionizes, and transforms
into the much more extended and more tenuous \x\ plasma.

\end{enumerate}

\acknowledgments This work is based on observations obtained with
\xmm\, an ESA science mission with instruments and contributions
directly funded by ESA Member States and the \usa\ (NASA). EB
acknowledges helpful discussions on \agn\ phenomenology with Ari
Laor. EB was supported by the Yigal-Alon Fellowship and by the GIF
Foundation under grant \#2028-1093.7/2001. The MSSL authors
acknowledge the support of PPARC.  SRON is supported financially
by NWO, the Netherlands Organization for Scientific Research. MS
was supported by NASA through \chandra\ Postdoctoral Fellowship
Award Number PF1-20016 issued by the \chandra\ \x\ Observatory
Center, which is operated by the Smithsonian Astrophysical
Observatory for and on behalf of NASA under contract NAS8-39073.
We thank an anonymous referee for a thorough and helpful report.

\clearpage
\bibliographystyle{apj}

\clearpage

\begin{deluxetable}{lccc}
%\tabletypesize{\scriptsize}
\tablecaption{Ions seen in absorption in the \rgs\ spectrum of
\ngc. \label{tbl-1}} \tablewidth{0pt} \tablehead{ \colhead{Ion} &
\colhead{Outflow Velocity [\kms]} & \colhead{$v_{turb}$
[\kms]~\tablenotemark{a}} & \colhead{$N_i$~[10$^{18}$ \cmsq]} }
\startdata
\ion{O}{8} & $-$550 $\pm$ 230 & 170 $\pm$ 50 & 4.0 $\pm$ 1.0 \\
\ion{O}{7} & $-$660 $\pm$ 230 & 170 $\pm$ 50 & 1.0 $\pm$ 0.3 \\
\ion{O}{6} & $-$800 $\pm$ 230 & 170 $\pm$ 50 & 0.01 $\pm$ 0.01 \\
\ion{O}{5} & $-$470 $\pm$ 230 & 170 $\pm$ 50 & 0.18 $\pm$ 0.03 \\
\ion{O}{4} & $-$770 $\pm$ 230 & 170 $\pm$ 50 & 0.08 $\pm$ 0.02 \\[0.3cm]
\ion{O}{3} & ---              & ---          & $\le$~0.005 \\
\ion{O}{2} \tablenotemark{b} & $\pm$ 230        & $kT_i$=0.1 eV \tablenotemark{c}& 0.004 $\pm$ 0.01 \\
\ion{O}{1} \tablenotemark{d} & $\pm$ 230        & $kT_i$=0.02 eV & 0.025 $\pm$ 0.035 \\[0.3cm]
\ion{O}{7} Gal. \tablenotemark{e} & z = 0.0018~$\pm$~0.0008 & 40;
$kT_i$=0.1~keV &
0.084 $\pm$ 0.03 \\
\enddata
\tablenotetext{a}{ Turbulent velocities of \ion{O}{4} - \ion{O}{8}
were tied together in the fit.}
\tablenotetext{b}{ Blended with \ion{O}{1} in our galaxy. }
\tablenotetext{c} { Ion temperature. }
\tablenotetext{d}{ Blended with \ion{Ar}{16} in \ngc. }
\tablenotetext{e}{ Tentative identification. }
\end{deluxetable}

\clearpage

\begin{deluxetable}{lccccc}
%\tabletypesize{\scriptsize}
\tablecaption{Oxygen emission lines in the \rgs\ spectrum of \ngc.
\label{tbl-2}} \tablewidth{0pt} \tablehead{ \colhead{Ion} &
\colhead{Observed} & \colhead{Velocity Shift}& \colhead{Velocity
Width}& \colhead{Line Flux}&
\colhead{Line Flux} \\
& Wavelength & [\kms]~\tablenotemark{a} & [\kms]~\tablenotemark{b}
&
[10$^{-5}$~cts~\sm~\cmsq] & [10$^{-5}$~cts~\sm~\cmsq] \\
& [\AA] & & & & \citep{kaspi02} } \startdata
\ion{O}{8} Ly$\alpha$& 19.173 & 330 $\pm$ 230 & 740 $\pm$ 140 &  4.5 $\pm$ 0.8 &  2.32 $\pm$ 0.74 \\
\ion{O}{7} $r$       & 21.833 & 290 $\pm$ 230 & 740 $\pm$ 140 &  7.8 $\pm$ 1.5 &  4.58 $\pm$ 2.38 \\
\ion{O}{7} $i$       & 22.034 & 290 $\pm$ 230 & 740 $\pm$ 140 &  4.1 $\pm$ 1.3 &  3.52 $\pm$ 1.82 \\
\ion{O}{7} $f$       & 22.335 & 290 $\pm$ 230 & 740 $\pm$ 140 & 11.5 $\pm$ 1.6 & 10.13 $\pm$ 2.74 \\
\enddata
\tablenotetext{a}{ In the system of \ngc\ and tied for all
\ion{O}{7} lines.}
\tablenotetext{b}{ Tied for all lines.}
\end{deluxetable}

\clearpage

\begin{deluxetable}{lccc}
%\tabletypesize{\scriptsize}
\tablecaption{Measured and derived parameters for the \x\
emitter/absorber in \ngc\ as traced by \ion{O}{7} \& \ion{O}{8}.
\label{tbl-3}}

\tablewidth{0pt} \tablehead{ \colhead{Parameter} &
\colhead{\ion{O}{7} ($f$)} & \colhead{\ion{O}{8} (Ly$\alpha$)} &
\colhead{\uv\ \citep {gabel03} } } \startdata
$\xi$ [erg \sm\ cm] & 30 & 30 & \\
$L$ [10$^{43}$ erg \sm] & 1.5 & 1.5 & \\
$N_i$ [10$^{18}$ \cmsq] & 1.0 $\pm$ 0.3 & 4.0 $\pm$ 1.0 & \\
$f_q$ & 0.14 & 0.57 & \\
$f_{q+1}$ & 0.57 & 0.29 & \\
$A_O$ & 5$\times$10$^{-4}$ & 5$\times$10$^{-4}$ & \\
$N_H$ [10$^{22}$ \cmsq] & 1.43 $\pm$ 0.43 & 1.40 $\pm$ 0.35 & \\
$d$ [Mpc] & 41.8 & 41.8 & \\
$F_{ji}$ [10$^{-5}~$~cts~\sm~\cmsq] & 11.5 $\pm$ 1.6 & 4.5 $\pm$ 0.8 & \\
$P_{ji}$ [10$^{-12}~$\sm~cm$^3$] & 3.9 & 5.7 & \\
$EM$ [10$^{64}$~\cmcu] & 2.2 $\pm$ 0.3 & 1.2 $\pm$ 0.3 & \\
&&& \\
\multicolumn{4}{c}{Localized Clouds Scenario  ($r_{max} - r_{min} = \Delta r \ll r$)} \\
$r$~[pc] & 16.3  $\pm$ 7.3 & 12.3  $\pm$ 4.9 & $\lesssim$~0.26 \\
$\Delta r$~[pc] & 9.9 $\pm$ 3.0 & 8.1 $\pm$ 2.1 & 3.3 $\times$10$^{-9}$ \\
$n_e$~[\cmcu] & 200 $\pm$ 180 & 340 $\pm$ 270 & 10$^{9}$ \\
&&& \\
$v_{out}$ [\kms] & $-$600 $\pm$ 150 &  $-$600 $\pm$ 150 & $-$550 - $-$1300\\
$M$ [10$^{37}$~g] & 1.0  $\pm$ 1.3 & 1.0 $\pm$ 1.1 & 1.8 $\times$10$^{-22}$ \\
$\dot{M}$ [10$^{33}$~g~yr$^{-1}$] & 0.6 $\pm$ 0.7 & 0.8 $\pm$ 0.8 &
3.4 $\times$10$^{-13}$ \\
$\Omega \cong (\Delta r / r)^2$ [str] & --- & --- & 1.6
$\times$10$^{-16}$ \\
&&& \\
\multicolumn{4}{c}{Ionization Cones Scenario  ($r_{max} \gg r_{min}$)} \\
$r_{min}$ [pc] & 9.4 $\pm$ 2.8 & 9.6 $\pm$ 2.4 & $\lesssim$~0.26 \\
$n_e (r_{min})$~[\cmcu] & 590 $\pm$ 350 & 560 $\pm$ 280 & 10$^{9}$ \\
$\Omega / 2\pi$ & 0.24 $\pm$ 0.08 & 0.13 $\pm$ 0.05 & 2.6 $\times$10$^{-17}$ \\
&&& \\
$v_{out}$ [\kms] & $-$600 $\pm$ 150 &  $-$600 $\pm$ 150 & $-$550 - $-$1300\\
$M/r_{max}$ [10$^{36}$ g pc$^{-1}$] & 8.4 $\pm$ 2.6 & 4.7 $\pm$ 1.8 & $M$~=~1.8
$\times$10$^{15}$~g \\
$\dot{M}$ [10$^{33}$~g~yr$^{-1}$] & 5.2 $\pm$ 2.0 & 2.7 $\pm$ 1.0 & 3.4$\times$10$^{-13}$ \\
\enddata
\end{deluxetable}

\clearpage
\begin{figure}
\plotone{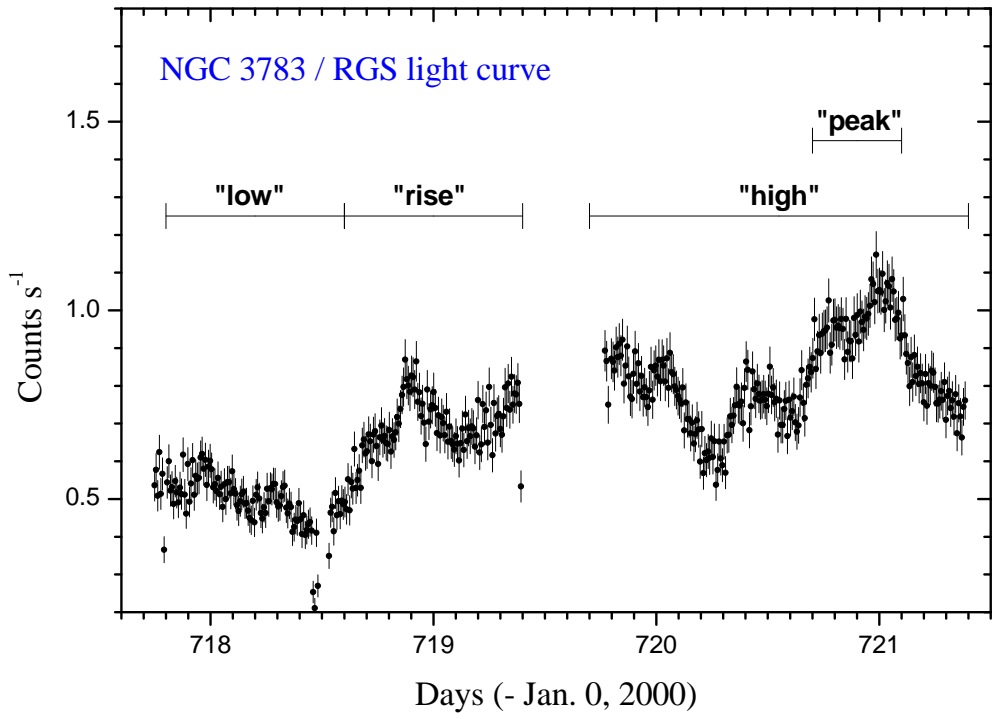} \caption{The \rgs\ light curve (\rgs\ 1 and 2
combined) during the present 280~ks observation in time bins of
approximately 10 minutes. Different segments of the observations
have been labeled. \label{fig1}}
\end{figure}

\clearpage
\begin{figure}
\epsscale{.90}
\plotone{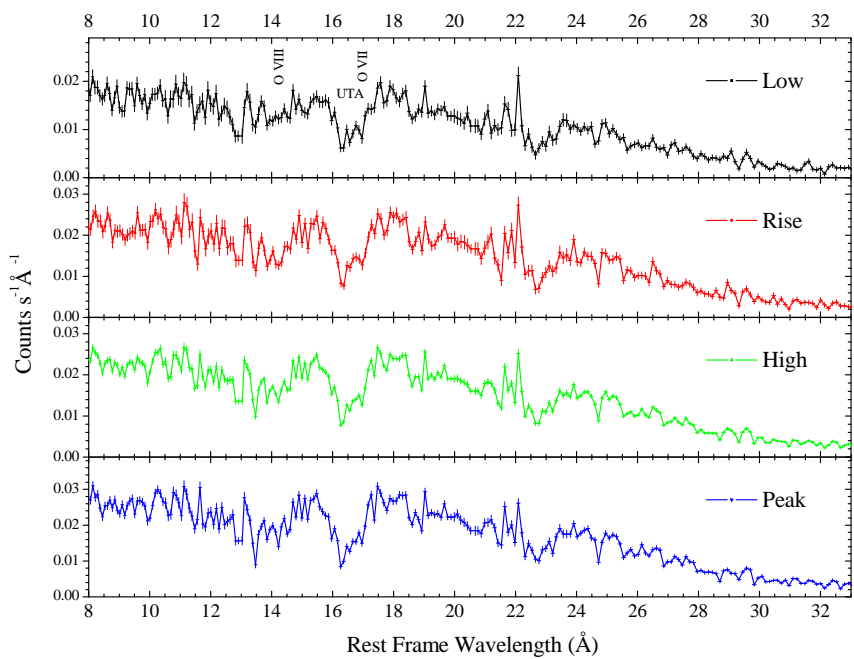} \caption{The \rgs\ spectra during
the four time segments defined in Fig.~1. The data are plotted in
bins of about 0.1~\AA\ to accentuate broad structure. Hence also
the small error bars. The lines between the data points are
plotted just to guide the eye. Apart from the different flux
levels, the shape of the absorption structure appears to be
remarkably similar in all four spectra. In particular, the broad
absorption features due to the Fe-M 2p-3d \uta\ (16~-17~\AA) and
due to the O-K edges (14.2 \& 16.8 \AA) do not appear to vary. The
quality of the data is not good enough to enable detection of
variability in the narrow lines. \label{fig2}}
\end{figure}

\clearpage
\begin{figure}
\plotone{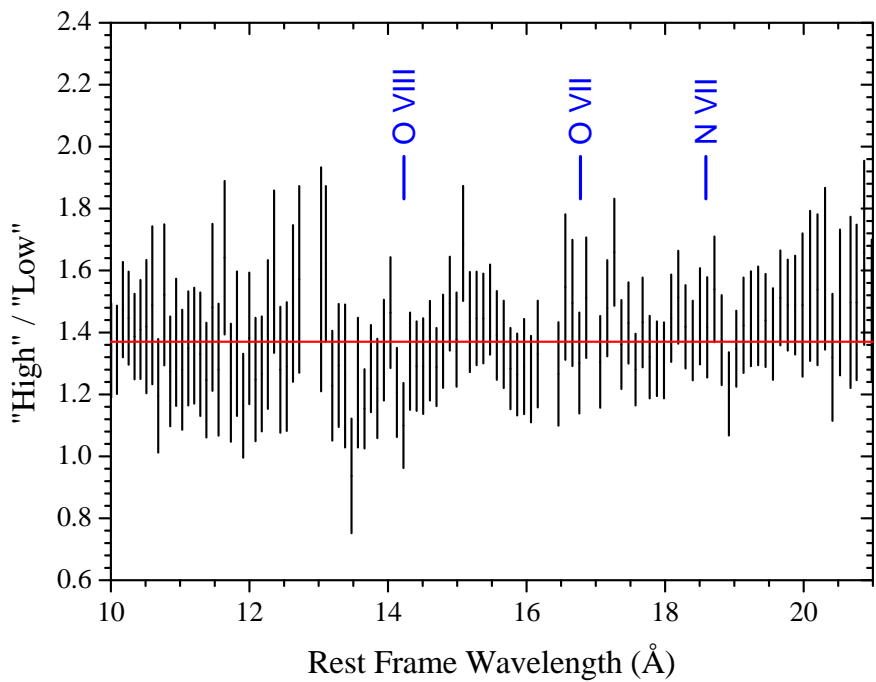} \caption{Ratio of the "high" phase and "low"
phase spectra rebinned to about 0.1~\AA. No appreciable broad
deviations from the mean (horizontal line) are found implying no
significant change in the photoelectric edges (marked above the
data). The gaps in the plot are instrumental. Note that even with
no change in absorption, the ratio around strong emission and
absorption lines (e.g., at 13.5, 14.2, 15, and 19~\AA) is not
expected to be perfectly flat, since deep absorption troughs
suffer from low S/N and emission lines do not cancel out in the
ratio. \label{fig3}}
\end{figure}

\clearpage
\begin{figure}
%\epsscale{.90}
\plotone{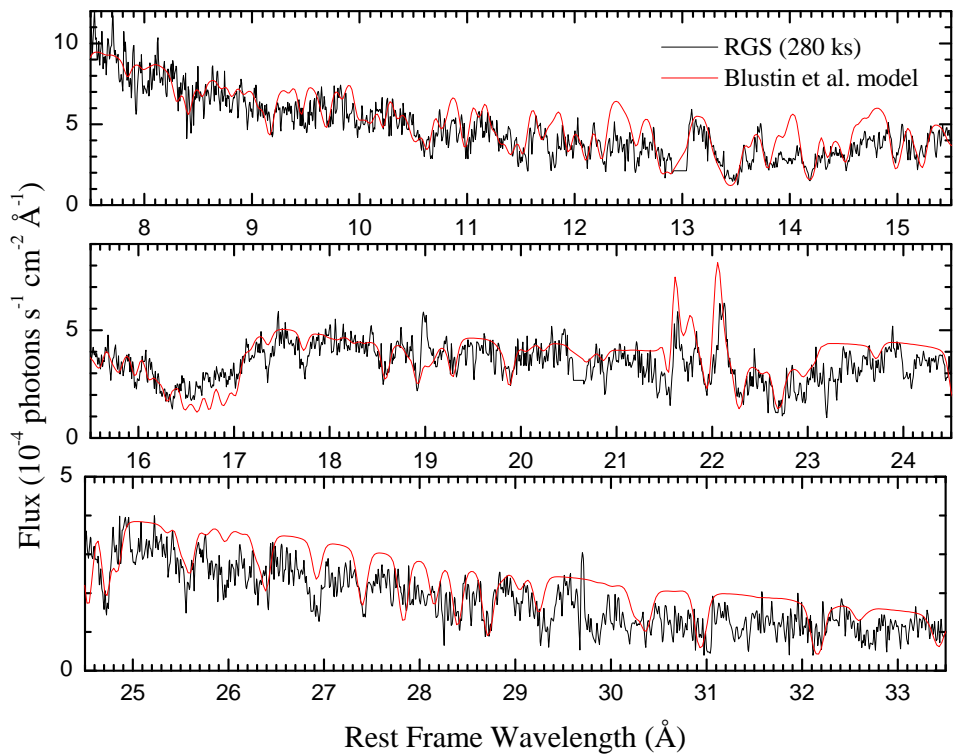} \caption{The total 280~ks \rgs\ spectrum of \ngc.
The slab model from \citet {blustin02}, which was fitted to the
first 40~ks RGS observation, is over-plotted with no adjustments,
including no re-normalization. The over all agreement is very
good, but noticeable discrepancies are seen in the details of the
O-K region between 14~and 24~\AA. \label{fig4}}
\end{figure}

\clearpage
\begin{figure}
\plotone{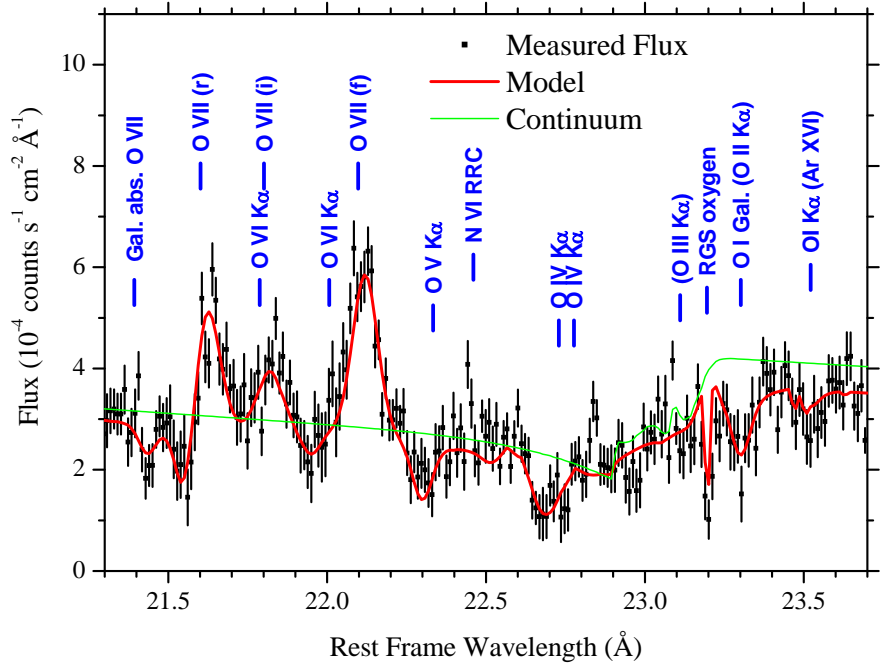} \caption{Spectrum of \ngc\ in the region of the
K$\alpha$ lines of \ion{O}{1} - \ion{O}{7} as obtained with a
280~ks \rgs\ observation. The thick curve represents the best-fit
model and the thin curve is the continuum model only with galactic
and instrumental absorption.  \label{fig5}}
\end{figure}

\clearpage
\begin{figure}
\plotone{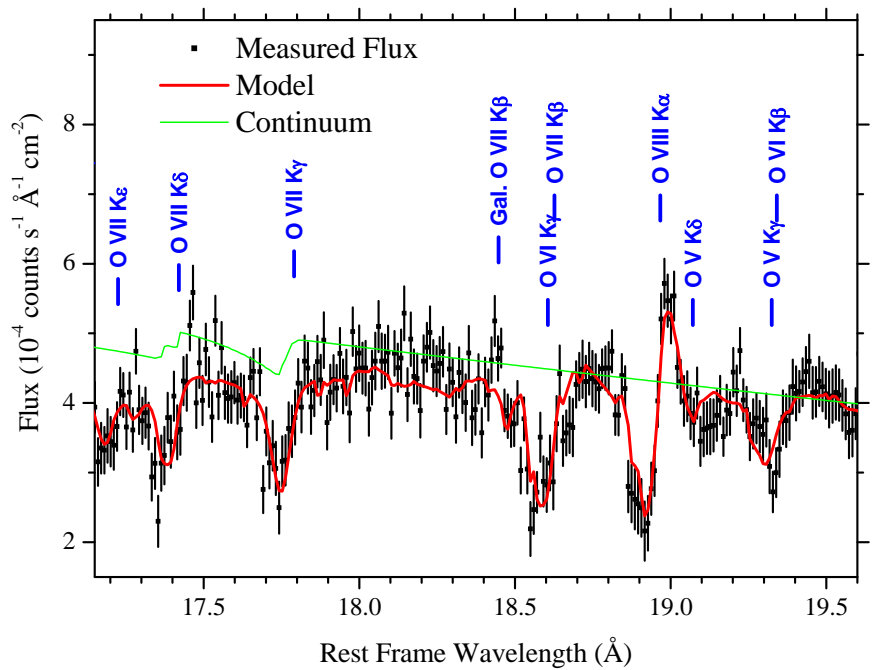} \caption{Spectrum of \ngc\ showing the region of
 \ion{O}{8} K$\alpha$ as well as higher order lines (K$\beta$,
 K$\gamma$, etc.) of lower charge states of O as obtained with a
 280~ks \rgs\ observation. The thick curve represents the
 best-fit model and the thin curve is the continuum model only with
 galactic and instrumental absorption.
\label{fig6}}
\end{figure}

\end{document}